\documentclass[prd,superscriptaddress,altaffilletter,amssymb,showpacs,nofootinbib,twocolumn,showkeys]{revtex4}
\usepackage[dvips]{graphicx} 
\usepackage[usenames]{color}
\usepackage[latin1]{inputenc}
\usepackage[english]{babel}
\usepackage{multirow}
\usepackage{rotating}
\usepackage{slashbox}

\usepackage{subfigure}
\newcommand{\be}{\begin{equation}}
\newcommand{\ee}{\end{equation}}
\newcommand{\bea}{\begin{eqnarray}}
\newcommand{\eea}{\end{eqnarray}}

\newcommand{\btheta}{\mbox{\boldmath $\theta$}}
\newcommand{\bsigma}{\mbox{\boldmath $\sigma$}}

\newcommand{\ben}{\begin{eqnarray}}
\newcommand{\een}{\end{eqnarray}}
\newcommand{\n}{\label}
\newcommand{\no}{\noindent}
\newcommand{\ga}{\gamma}

\newcommand{\om}{\Omega}
\newcommand{\bi}[1]{\mbox{\boldmath$#1$}} 

\begin{document}
\title{DBI models for the unification of dark matter and dark energy}
\author{Luis P. Chimento}
\email{chimento@df.uba.ar}
\affiliation{Departamento de F\'isica,
Universidad de Buenos Aires, 1428 Buenos Aires, Argentina}
\author{Ruth Lazkoz}\email{ruth.lazkoz@ehu.es}
\affiliation{Fisika Teorikoa, Zientzia eta Teknologia Fakultatea, Euskal Herriko Unibertsitatea, 644 Posta Kutxatila, 48080 Bilbao, Spain}
\author{Irene Sendra}\email{irene_sendra@ehu.es}
\affiliation{Fisika Teorikoa, Zientzia eta Teknologia Fakultatea, Euskal Herriko Unibertsitatea, 644 Posta Kutxatila, 48080 Bilbao, Spain}
\date{\today}

\begin{abstract}
We propose a model based on a DBI action for the unification of dark matter and dark energy. This is supported by the results of the study of its background behavior at early and late times, and reinforced by the analysis of the evolution of  perturbations. We also perform a Bayesian analysis to set observational constraints on the parameters of the model using type Ia SN, CMB shift and BAO data. Finally, to complete the study we investigate its kinematics aspects, such as the effective equation of state parameter, acceleration parameter and transition redshift. Particularizing those parameters for the best fit one appreciates that an  effective  phantom is preferred.
\end{abstract}
\maketitle



\section{Introduction}
Unified models of the two main components of the universe, dark energy and dark matter, represent an interesting
option for explaining the substantial evidence of the current acceleration of the universe. On the one hand no
observational direct evidence of either of them is available, so it might well be the case that they do simply
not exist and there is a bolder explanation for the effects we attribute to them (perhaps extra dimensions are the
answer). On the other hand, if one
believes these two entities really fill our universe (in a huge joint proportion as compared to the other components)
it remains to discover what their nature is. In a way, finding out that they happen to be manifestations of the same
fluid would at least simplify the problem in the sense one should have to care about the fundamentals of a single fluid.
At worst, if investigations along these lines were able to refute this idea that the two components are just one, we would
at least be able to face the future in the confidence that dark energy and dark matter can be treated separately.

Following the tradition of trying to find an interconnection between the world of Particle Physics and Cosmology,
it is customary to try and view unified dark energy models as scalar field scenarios. One possibility is to explore
the evolutions contained in a given scalar field model, this is actually the approach of this paper. We consider a scalar field
setup and by fixing some of its degrees of freedom we obtain an expansionary cosmology which mimics a dark matter dominated
background at early times and a dark energy dominated one at the late stages of its history. On the other hand, a popular
procedure to find a scalar fields based motivation for a given evolution is to start from a given equation of state and then ``reconstruct'' the corresponding Lagrangian (by specifying its kinetic term and potential). This widely followed approach
has its caveats, however, as in general the scalar field model one ends up with is in fact a richer scenario, and contains
other evolutions that the original seed. Leaving aside these remarks, the route of scalar fields toward unified dark energy scenarios
may offer interesting possibilities and our efforts in this paper go in this direction.

It has been put suggested that acceleration is cosmological settings  might be the manifestation
of non-perturbative features of
of some string theory versions  \cite{Quevedo:2002fc}. This idea has gathered quite a lot of attention as it could provide
an explanation to early time acceleration, that is, inflation,  (see \cite{Dvali:1998pa,Leigh:1989jq,Kachru:2003sx,Brodie:2003qv,Cline:2006hu,Kallosh:2007ig,HenryTye:2006uv,
Silverstein:2003hf,Alishahiha:2004eh,Chen:2004hua,Chen:2005ad,
Kecskemeti:2006cg,Lidsey:2006ia,Baumann:2006cd,Spalinski:2007qy,Krause:2007jr,Huang:2007hh,Bean:2007eh,
Gmeiner:2007uw,Spalinski:2007dv,Spalinski:2007kt}
and references therein for regular papers and \cite{Peiris:2007gz} for a review). According to this description, the
inflation could be a mode accounting for  the
position of a D-brane with three spatial dimensions rambling (radially) in a ten-dimensional space-time with
a warped metric. This interpretation seems to have the virtue, among others, that it would allow inflation to
proceed with much steeper potentials than in the standard
weakly coupled slow roll inflation model.

So, a somewhat natural question to ask is this one: could a DBI model be responsible for the acceleration we observe at present?
 Moreover, given the similarity of the DBI Lagrangian to that of the most popular unified dark energy model, the Chaplygin gas, could it also offer
a satisfactory and perhaps even more suitable alternative for the unification of dark matter and dark energy? The investigations we report in this paper show that is indeed the case . We construct a purely kinetic DBI model for the joint description of the two main components of the universe with the bonus that it also displays late-time phantom behavior without resorting to true phantom dark energy at all. 

Any conclusions about the capability of our model to represent a solid alternative to other dark energy scenarios
must ideally be reached from both the theoretical and   the observational
perspectives. To that end  first we carry out the computation and interpretation of the linear gauge invariant perturbations of the model. After that, we perform a thorough analysis of this novel unified dark sector scenario using geometric means: specifically we use
the SNIa, the BAO, and the CMB shift test. These combination of tests allows to take into account 
the early, mid and late time behavior of our model, which is expected to have its own features
as compared to models in which dark energy and dark matter are different components.  Our analysis
is performed in the Bayesian spirit and it allows us to identify the best fit and errors, and to complete
the information obtained with a computation of the evidences on different ranges of the parameters
and constraints on kinematical quantities of interest. 
\section{The model}
Our scenario is that of a four-dimensional spatially flat FRW spacetime filled with a non-canonical scalar field of DBI type. Using the customary perfect fluid interpretation we set
\begin{eqnarray}
 \rho=\frac{\ga-1}{f}+V(\phi)\label{rhodef},\label{rho}\\
p=\frac{\ga-1}{\ga f}-V(\phi)\label{pdef},
\end{eqnarray}
with
\begin{equation}
\n{g}
\gamma=\frac{1}{\sqrt{1- f(\phi)\dot\phi^2}},
\end{equation}
where, in principle, $f$ and $V$ are arbitrary functions.  Usage of  the symbol $\gamma$ was originally motivated by its analogy to the Lorentz factor of Special Relativity, given that $\sqrt{f(\phi)}\dot\phi$ is interpreted as the proper velocity of the brane \cite{Silverstein:2003hf}. 

Assuming for the above fluid a barotropic equation of state  of the form $p=(\Gamma-1)\rho$, we get 
\be
\n{G}
\Gamma=-\frac{2\dot H}{3H^2}=\displaystyle\frac{\gamma\dot\phi^2}{\rho},
\ee
and the conservation equation reads
\begin{eqnarray}
\dot\rho+3H\Gamma\rho=0. \n{con}
\end{eqnarray}

In this paper we explore the case in which both $f=f_0$ are $V=V_0$ constants.
The goal is to obtain a purely kinetic model as in other unified dark sector models 
\cite{Kamenshchik:2001cp,Bento:2002ps,Bilic2002,Scherrer:2004au,Chimento:2004jm}, so that the field $\phi$ depends solely on the scale factor; and as a consequence the same holds for the effective
pressure and energy density. To that end we insert
\be
\n{r}
\rho=\frac{\ga-1}{f_0}+V_0,
\ee
into the conservation equation (\ref{con}) and obtain 
\be
\n{.g}
\dot\ga+3f_0H\ga\dot\phi^2=0.
\ee
Upon replacement of the derivative $\ga$,
\be
\n{.gg}
\dot\ga=f_0\ga^3\dot\phi\ddot\phi.
\ee
and using the Eqs. (\ref{g}), (\ref{.g}) we arrive at
\be
\frac{\ddot\phi}{\dot\phi}+3H+\frac{\dot\ga}{\ga}=0,
\ee
which has the following first integral:
\be
\dot\phi=\frac{c}{\ga}\left(\frac{a_0}{a}\right)^3.
\ee
Thus, it is possible to write 
\be
\ga^2=1+c^2f_0\left(\frac{a_0}{a}\right)^6.\label{gsq}
\ee
with  $c$ an arbitrary integration constant and  $a_0$ the value of the scale factor today, which we fix as $a_0=1$.

On the one hand we have accomplished our goal of obtaining a DBI model which is purely kinetic throughout the evolution. On the other hand, the behavior of the model obtained is quite appealing.  Using (\ref{gsq}) it can be seen that at very late times, i.e. in the regime $a\gg a_0$, one has  $\rho\sim V_0$, whereas
at early epochs, i.e. for $a\ll a_0$, one has $\rho\sim 1/a^3$ instead. Thus, synthesizing, the solution  found interpolates between a dust and a de Sitter model.
Of course, for a positive $f_0V_0$ it is easy to see $H>0$, so the evolution is indeed expansionary.

This new model represents an alternative description for the unification of dark matter and dark energy, and as popular models of this sort, it can be linked to a non-canonical scalar field model obtained from the following Lagrangian \cite{Spalinski2007}
\be
\mathcal{L}(X)=-\frac{1}{f_0}\left(\sqrt{1-f_0X}-1\right)-V_0.
\ee
with $X=\dot{\phi}^2$.Interestingly, the background evolution of our model is equivalent to that of a model filled with two separately conserved  components: a conventional Chaplygin gas (Cg) and a cosmological constant (cc). There is, however, an important difference with respect to our model, as the equivalence is limited to zeroth order cosmology, i.e. to the background evolution. At the level of perturbations, differences are bound to appear, as in our model this evolution is realized with a single fluid. It is actually possible to derive its equation of state
\be
p=\frac{1}{f_0}\left(1-\frac{1}{\rho f_0-V_0f_0+1}\right)-V_0
\ee
and to draw from it the same conclusion about the asymptotic behavior of the model.
In contrast, if one would like to stick to a description in terms of  two components (Cg+cc) one should demand there is an interaction among them so they act like a single fluid,
and given the degree of arbitrariness involved in the use of interaction terms, we will rather focus on the single fluid description provided by the DBI picture. 

Nevertheless, one must keep in mind the possibility out of the scope of this paper that the universe is made of 
two fluids, the DBI one plus cosmic dust, which ultimately should be confronted with the model we present here. 
This possibility has been recently explored from an asymptotic behaviour perspective in \cite{Martin2008}.

\section{Linear perturbations}
In the synchronous gauge the line element is given by: 
\be
\n{ds}
ds^2 = a^2(\tau)[-d\tau^2 + (\delta_{ij} + h_{ij})dx^idx^j], 
\ee
where the comoving coordinates are related to the proper time $t$ and position ${\bi r}$ by $d\tau=dt/a$, $d{\bi x}=d{\bi r}/a$, and $h_{ij}$ is the metric perturbation.
The scalar mode of $h_{ij}$ is described by the two fields $h({\bi k},\tau)$ and $\eta(\bi k,\tau)$ in the Fourier space: 
\be
\n{def}
h_{ij}(\bi x,\tau)=\int d^3 k e^{i\bi k{\bi\cdot}\bi x}\left[\hat{\bi k_i }\hat{\bi k_j} h + (\hat{\bi k_i }\hat{\bi k_j} -\frac{1}{3}\delta_{ij})6\eta \right], 
\ee
with ${\bi k} = k \hat{\bi k}$. The Einstein equations to linear order \footnote{It has been pointed out that linear perturbations
may not be sufficient to treat unified dark sector models and a methods to do so 
has been proposed in \cite{Bilic2008}
}
in
k-space, expressed in terms of $h$ and $\eta$, are given by the following
four equations \cite{Ma1995}:
\ben                                                                  
\n{a1}
&&k^2\eta - \frac{1}{2}\frac{a'}{a}h'= 4\pi G a^2 \delta T^0_0, \\
\n{a2}
&&k^2 \eta' = 4\pi G a^2 (\rho + p)\theta,   \\
\n{a3}
&&h'' + 2\frac{a'}{a}h' -2k^2\eta = -8\pi G a^2 \delta T^i_i,  \\  
\nonumber
&&h''+6\eta''+2\frac{a'}{a}(h'+6\eta') -2k^2\eta =\\
\n{a4}
&&-24\pi G a^2 (\rho + p)\sigma.
\een
Here, the quantities $\theta $ and $\sigma$ are defined as $(\rho + p)\theta = i k^j \delta T^0_j$, $(\rho + p)\sigma = -(\bi k_i\bi k_j - \delta_{ij}/3)\Sigma^i_j$ and $\Sigma^i_j = T^i_j - \delta^i_j T^k_k/3$ denotes the traceless component of the tensor $T^i_j$. In addition,  $\theta $ is the divergence of the fluid's velocity $\theta = i k^jv_j$ and $'$ means $d/d\tau$.

Let us consider a fluid moving with a small coordinate velocity $v^i = dx^i/d\tau$, then, $v^i$ can be treated as a perturbation of the same order as the energy density, pressure and metric perturbations. Hence, to linear order in the perturbations, the energy-momentum tensor, with vanishing anisotropic shear perturbation $\Sigma^i_j$, is given by
\ben
\n{tpert1}
T^0_0 = - (\rho + \delta\rho), \\
\n{tpert2}
T^0_i =  (\rho + p)v_i = - T^i_0, \\
\n{tpert3}
T^i_j =  (p + \delta p)\delta^i_j.
\een

For a fluid with equation of state $p=w\rho$, the perturbed part of energy-momentum conservation equations $T^{\mu\nu}_{\phantom{\mu};\mu} = 0$ in the k-space leads to the equations
\ben
\n{pertd}
&&\delta' = -(1+w)\left(\theta + \frac{h'}{2}\right) - 3\mathcal H \left(\frac{\delta p}{\delta \rho} - w\right)\delta,   \\
\n{tet}
&&\theta' = - \mathcal H (1 -3w)\theta - \frac{w'}{1+w}\theta
+ \frac{\delta p/\delta \rho}{1 + w}k^2\delta,
\een
\no where $\delta = \delta\rho/\rho $ and $\mathcal H = a'/a=a H=\dot a$. 
Assuming strictly adiabatic contributions to the perturbations, the speed of sound for the fluid is 
\be
\n{cs}
c_s^2=\frac{\delta p}{\delta\rho}=\frac{\dot p}{\dot\rho}=\frac{1}{\ga^2}=\frac{a^6}{c^2f_0+a^6}, 
\ee
and the time variation of $w$ is 
\be
\n{w'}
w'=-3\mathcal H(1+w)(c_s^2-w).
\ee
Hence, inserting these last two equations in (\ref{pertd}) and (\ref{tet}), they become
\ben
\n{pertd1}
&&\delta' = -(1+w)\left(\theta + \frac{h'}{2}\right) - 3\mathcal H (c_s^2 - w)\delta,   \\
\n{tet1}
&&\theta' = - \mathcal H (1 -3c_s^2)\theta+\frac{c_s^2}{1 + w}k^2\delta.
\een
Besides, using equations (\ref{a1}), (\ref{a3}), (\ref{tpert1}) and (\ref{tpert3}) we arrive at 
\be
\n{h}
h'' + \mathcal H h' + 3\mathcal H^2(1 +3c_s^2)\delta = 0.
\ee

At early time, when the overall fluid has $w\approx 0 $, the effective fluid perturbations evolve similar to those of ordinary dust with $\dot\theta=\theta=0$, $a\sim t^{2/3}$ and from Eqs. (\ref{pertd},\ref{h}) we obtain
\be
\n{early1}
\delta'' + \mathcal H \dot\delta - \frac{3}{2}\mathcal H ^2\delta = 0
\ee 
and $\delta=c_1 t^{-1}+c_2 t^{2/3}$, where $c_1$ and $c_2$ are arbitrary integration constants. In this dust dominated era the perturbation grows as $\delta\approx a$ showing an initial unstable phase, compatible with the observation that the primordial universe would have tiny perturbations which seeded the formation of structures in the  universe. Conclusions about the clustering capabilities
of other cosmic settings with DBI fluids have been studied in \cite{Bertacca2008}.

At late times, we are interested to find the evolution of the linear scalar perturbations for any mode $k$. To this end we write the second order differential equation for the density perturbation $\delta$, see \cite{Bean2003}
\ben
\nonumber
&&\delta''+[1+6(c_s^2-w)]\mathcal{H}\delta'+\left[9(c_s^2-w)^2\mathcal{H}^2\right.\\
\nonumber
&&\left.+3(c_s^{2'}-w')\mathcal{H}+3(c_s^2-w)(\mathcal{H}'+\mathcal{H}^2)+c_s^2k^2\right.\\
&&\left.-\frac{3}{2}(1+3c_s^2)(1+w)\mathcal{H}^2\right]\delta=-3c_s^2(1+w)\mathcal{H}\theta.
\n{delt1}
\een
Taking into account that in the late time regime the scale factor behaves as $a\propto e^{\sqrt{\frac{V_0}{3}}t}$ we can calculate $\mathcal{H}$
\ben
\n{mh}
\mathcal{H}=a'/a=\dot a\propto a.
\een
Considering the expression of $\ga$ given by the Eq. (\ref{gsq}) one obtains the late-time expansion in terms of $1/a$ for $\rho$, $p$, $w$, $c_s^2$, $w'$ and $\mathcal{H}$. In this way, replacing these expansions in Eqs. (\ref{tet1},\ref{delt1}) and keeping only the most significant terms one gets
\ben
\nonumber
&&\delta''+13\mathcal{H}\delta'+\left[9(c_s^2-w)^2\mathcal{H}^2+2(c_s^{2}-w)\mathcal{H}^2
+c_s^2k^2\right]\delta\\
&&=-3c_s^2(1+w)\mathcal{H}\theta\n{delt2}\\
&&\theta' =2\mathcal H \theta+\frac{V_0k^2a^6}{c^2}\delta.
\n{t2}
\een

From Eqs. (\ref{delt2}) and (\ref{t2}) the  evolution of the perturbation becomes mode dependent with the $k^2/\mathcal{H}^2$ term, and for low energy modes their solutions can be obtained assuming a power law dependence of the perturbations with the scale factor, $\delta\propto a^n$ and $\theta\propto a^s$. In this case the approximate solutions are given by
\ben
\n{st}
&&\theta\approx\theta_0 a^{2}\\
\n{sd}
&&\delta\approx \frac{\delta_1}{a^{4}}+\frac{\delta_2}{a^{10}}+\frac{\theta_1}{a^{5}},
\een
where $\theta_0$, $\delta_1$ and $\delta_2$ are integration constants while $\theta_1$ is a function of $\theta_0$, $c$ and $V_0$. This shows that the coupling to $\theta$ in Eq. (\ref{delt1}) can be neglected for all scales we are interested on. Also we find that the energy density perturbation decreases for large cosmological times for modes satisfying the condition $k^2/\mathcal{H}^2\ll 1$. For high energy modes, $k^2/\mathcal{H}^2\gg 1$, 
Eq. (\ref{delt2}) is like the
equation of motion of a dissipative mechanical system. This resemblance emerges using the analogy with the classical potential problem
\be
\n{em}
\frac{d}{d\tau}\left[\frac{\delta'^2}{2}+\mathcal{V}(\delta)\right]=-13\mathcal{H}\delta'^2,
\ee
where 
\be
\n{V}
\mathcal{V}(\delta)=\frac{k^2\delta^2}{2},
\ee
As for any mode $k$ the potential $\mathcal{V}$ has a minimum at $\delta=0$,  the function inside the square bracket in Eq. (\ref{em}) is a Liapunov function and the perturbation decreases asymptotically reaching $\delta=0$ in the limit $t\to\infty$.

\section{Observational constraints}
In this section we will set constraints on the parameters of the model from a Bayesian perspective. Our analysis will use geometrical tests: the SN type Ia luminosity test, the CMB shift test \cite{Wang2007,Komatsu2008}, and the BAO test \cite{Percival2007}. As these two  last tests involve early universe quantities (the sound horizon at decoupling and dragging epochs), one must
consider a slightly more general setup and include radiation, which must to be conserved independently from the DBI fluid. This way, 
the Friedmann equation turns out to be
\be
\n{00}
3H^2=f_0^{-1}\left({\sqrt{1+c^2f_0\left(\frac{a_0}{a}\right)^6}-1}\right)+V_0+\rho_{r0}\left(\frac{a_0}{a}\right)^4.
\ee
In terms of the fractional energy densities and the redshift we have
$$
\frac{H^2}{H_0^2}=\sqrt{\Omega_f^2+\Omega_c^2(1+z)^6}+\Omega_{\Lambda}+\Omega_r(1+z)^4,
\label{eq:fri}
$$
where
\begin{eqnarray}
&&\om_f=\frac{1}{3H_0^2f_0},\\
&&\Omega_{\Lambda}=\frac{f_0V_0-1}{3H_0^2f_0},\\
&&\Omega_{r}=\frac{\rho_{r0}}{3H_0^2},
\end{eqnarray}
The latter are subject to the normalization condition 
\be
\sqrt{\Omega_f^2+\Omega_c^2}+\Omega_{\Lambda}+\Omega_r=1.
\ee 
In addition, the CMB and BAO tests require that we identify a combination of parameters which behaves effectively as $\Omega_m$ in the high energy regime. In our case this mimicry is played  by $\Omega_c$.

As this paper is a first approach to this model, we are setting constraints only on $\Omega_c$ and $\Omega_f$. In contrast, we fix a prior for $\Omega_b$, taking the WMAP 5-year best fit, $\Omega_b=0.0432$. Using the tests mentioned before in the framework of Bayesian statistics, one should minimize the corresponding $\chi^2$ function in order to obtain $\Omega_c$ and $\Omega_f$, see Appendix \ref{estsec}.

\begin{figure*}
\subfigure[\ UNION]{
\includegraphics[width=0.3\textwidth]{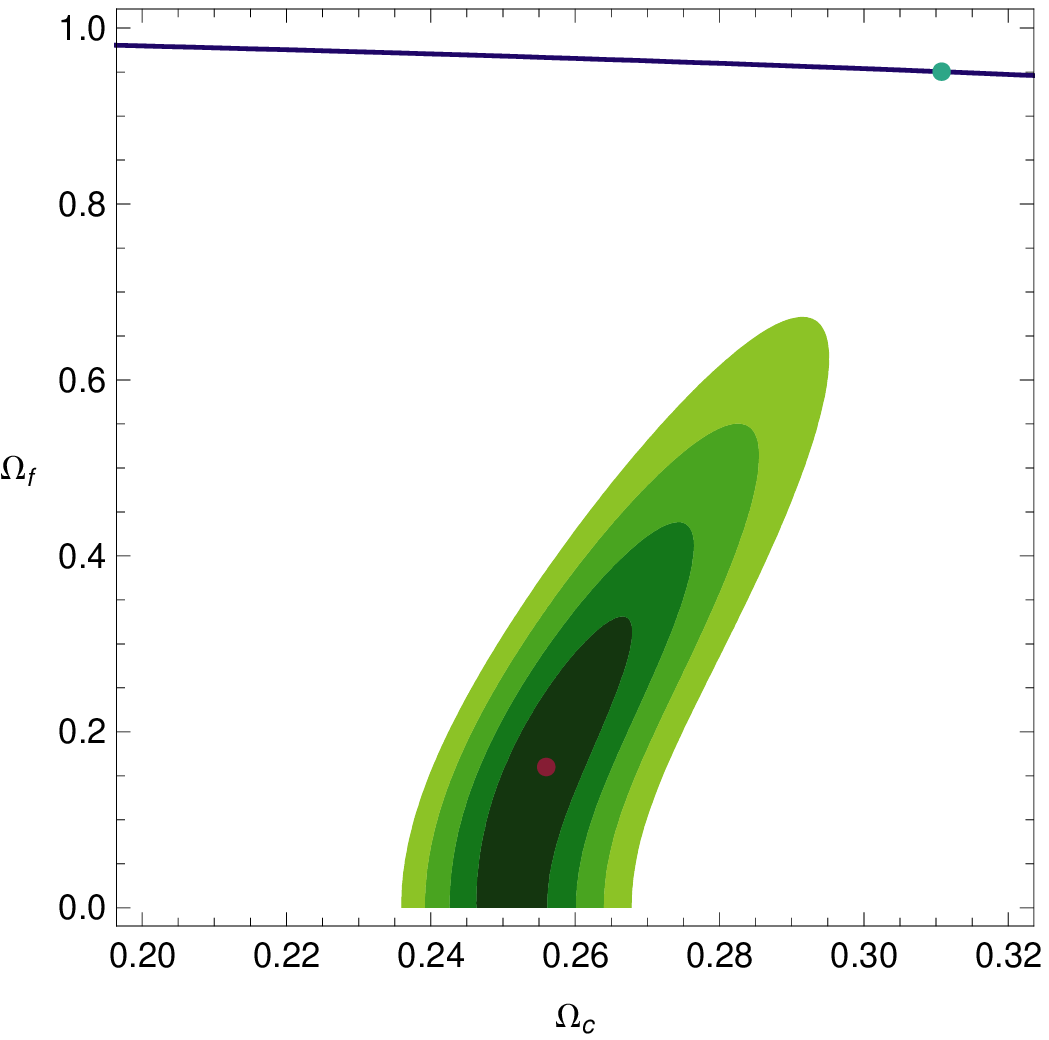}
\label{fig:CU}}
\subfigure[\ ESSENCE]{
\includegraphics[width=0.3\textwidth]{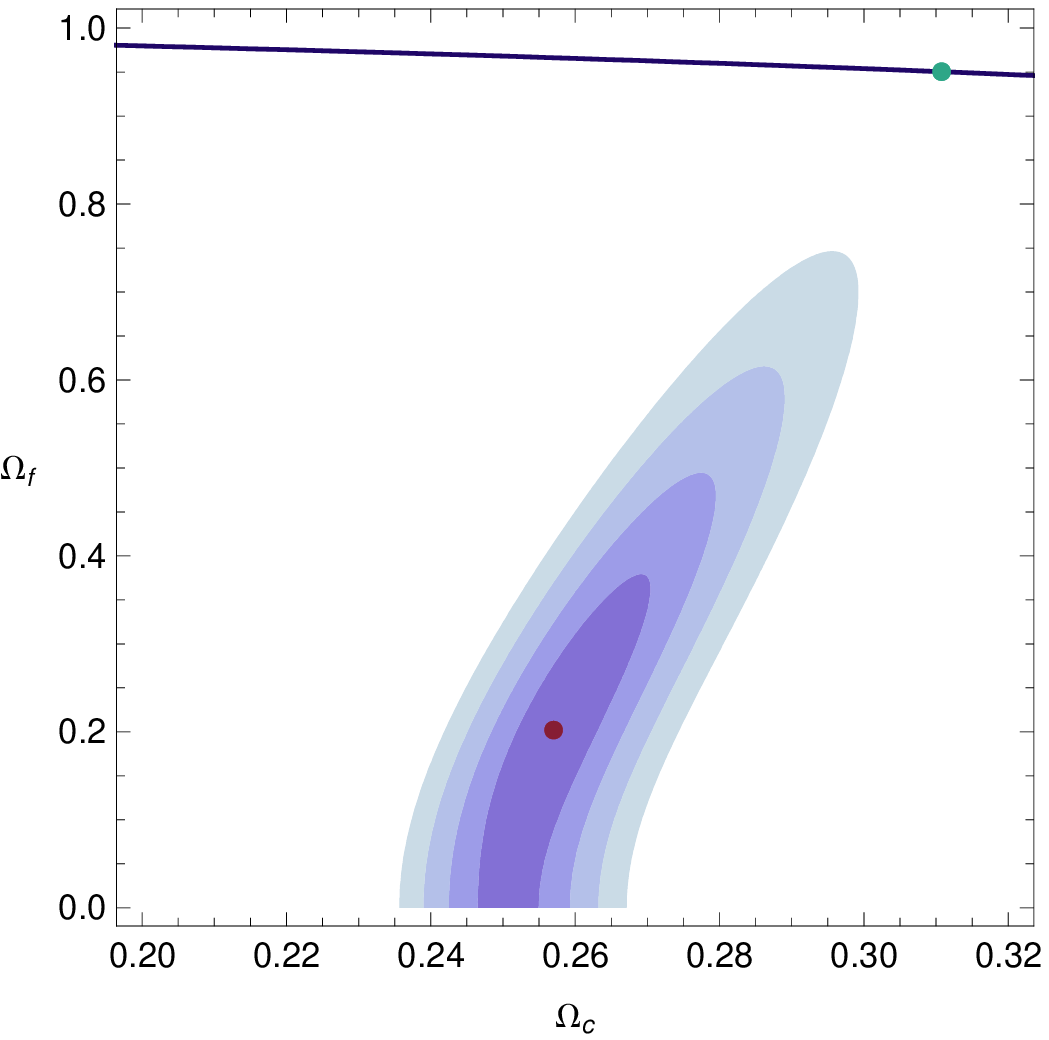}
\label{fig:CE}}
\caption{\label{fig:contours}Credible intervals from the combination of SN+CMB+BAO observations for two different SN compilation samples.}
\end{figure*}

We have used two different compilations for SNIa data: ESSENCE \cite{Krisciunas2008,Davis2007}, which combines the first results of the survey \cite{Wood-Vasey2007} with the results of Riess et al. detected by HST \cite{Riess2007} and UNION \cite{Kowalski2008}, a vast sample which brings together 414 SN from 13 independent datasets: recent samples (SLS, ESSENCE), old datasets and distant supernovae from HST. In the case of the UNION sample, the best values obtained are $\Omega_c=0.256^{+0.012}_{-0.010}$, $\Omega_f=0.160^{+0.171}_{-0.160}$ and for the ESSENCE sample $\Omega_c=0.257^{+0.013}_{-0.011}$ and $\Omega_f=0.202^{+0.177}_{-0.202}$ with the corresponding 68.30\% uncertainties. The lines in the upper sides of the plots in Fig.\ref{fig:contours} represent the locations on the parameter space which correspond to Chaplygin gas cases and the points in each of the lines indicate the case with the lowest $\chi^2$ value. From visual inspection one can infer that the Chaplygin !
 gas is rejected by our model. In contrast, LCDM (the $\Omega_f=0$ locations)  is not significantly excluded, as for a certain range of $\Omega_m$, LCDM cases lie in the 68.30\% likelihood credible interval. All in all our model provides better fits.

\begin{figure*}
\subfigure[\ UNION]{
\includegraphics[width=0.3\textwidth]{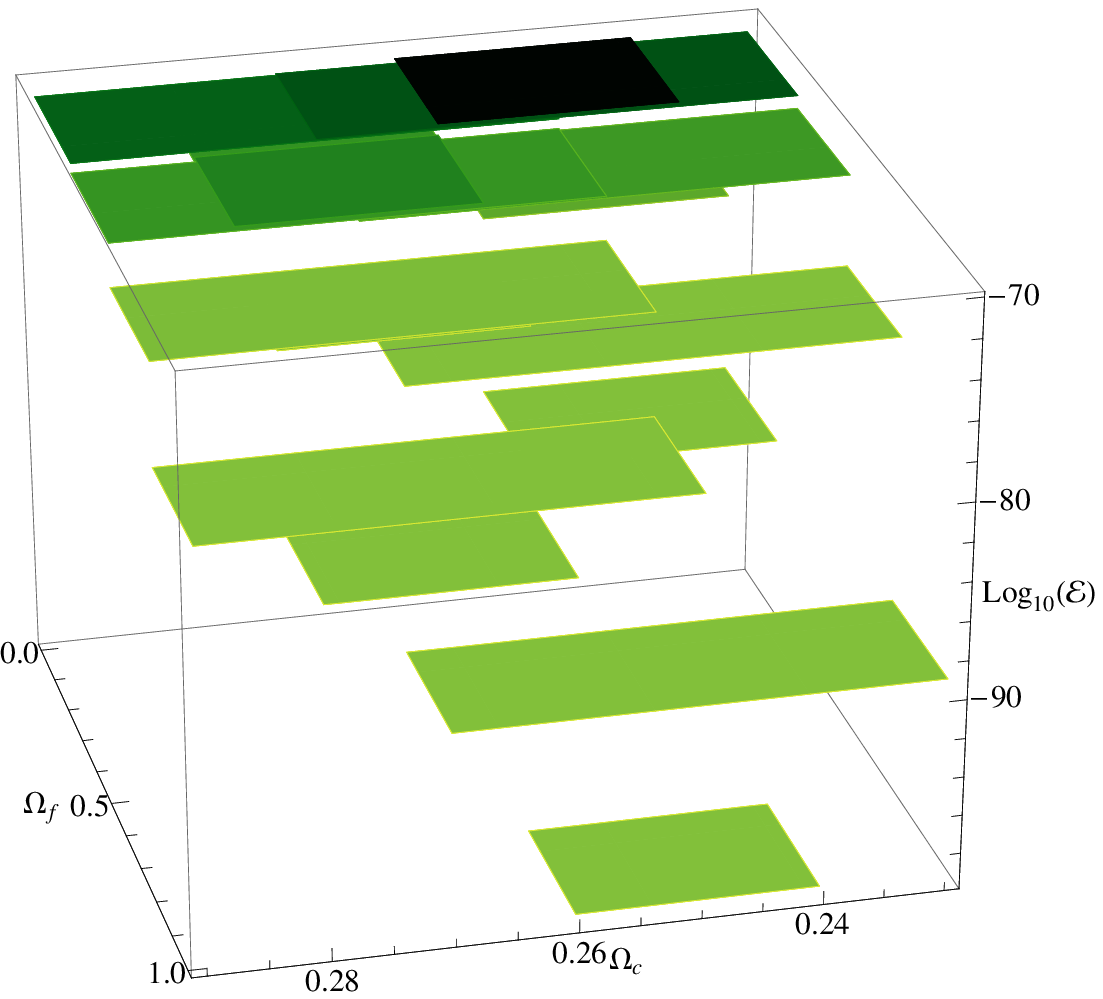}
\label{fig:3Devidence_union}}
\subfigure[\ ESSENCE]{
\includegraphics[width=0.3\textwidth]{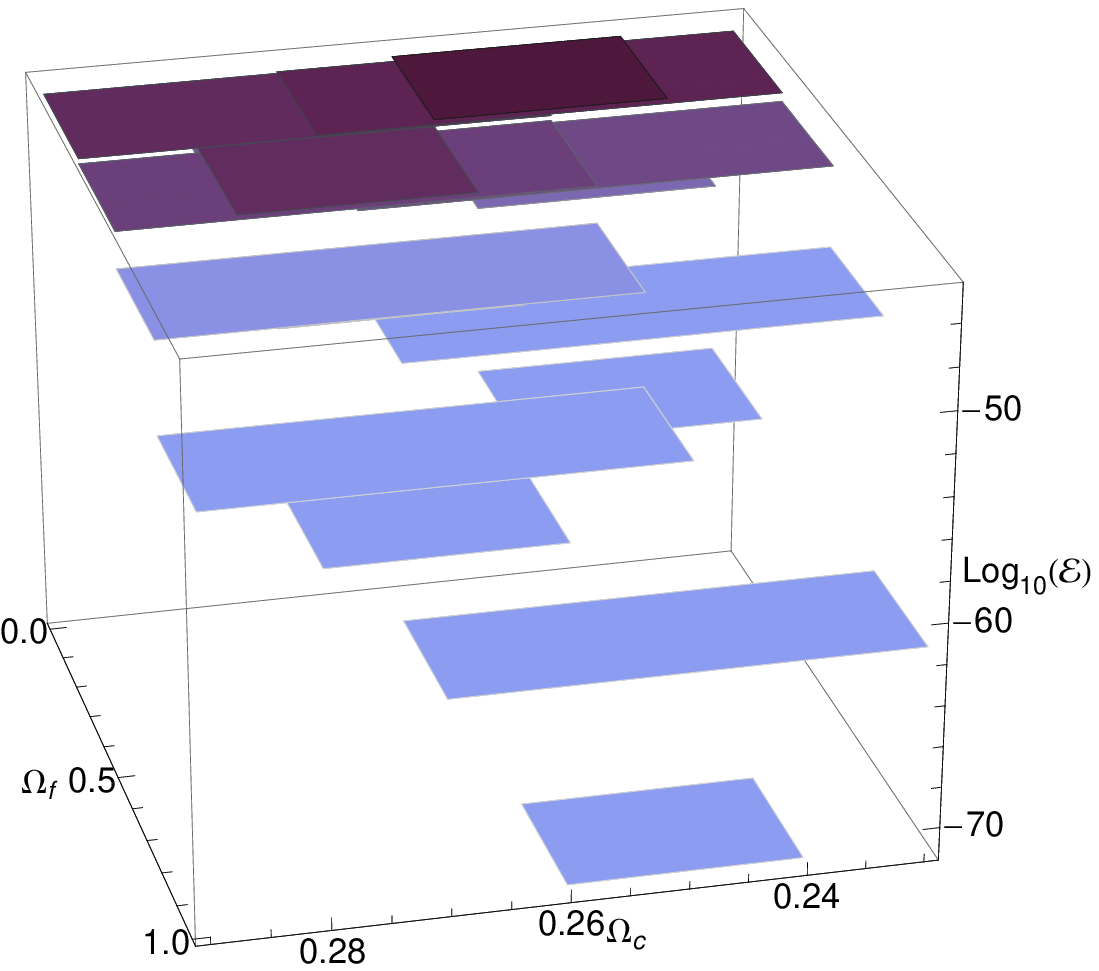}
\label{fig:3Devidence_essence}}
\caption{3D representation of Bayesian evidences for our unified dark energy DBI model from the combination of SN+CMB+BAO observations for two different SN compilation samples.\label{fig:3D_evidences}}
\end{figure*}

With the aim of compensating for the arbitrariness in the choice of priors, we explore different priors on $\Omega_c$ and $\Omega_f$. In the case of $\Omega_c$ we have the guidance of all the literature of constraints on dark energy constraints which more or less suggests preferred regions. To take advantage of this we explore four priors of different lengths, all centered at the value $\Omega_c=0.25$. In contrast, to illustrate the effect of changing the prior on $\Omega_f$, which is a new parameter on which we have no previous clues, we divide the physically allowed region $\Omega_f \in [0.00,1.0]$ into four equal intervals. From Fig. \ref{fig:3D_evidences} and Tab. \ref{tab:evidences} one can conclude that among the priors considered, the region $\Omega_f \in [0.00,0.25]$, $\Omega_c \in [0.24,0.26]$ gives the best constraints for the parameters.

\begin{table*}
\begin{center}
\subtable[\ UNION]{
\begin{tabular}{|c|c|c|c|c|}
\hline
\backslashbox{\textbf{Prior on $\Omega_f$}}{\textbf{Prior on $\Omega_c$}}& ~$\Omega_c \in [0.24,0.26]$ ~ & ~$\Omega_c \in [0.23,0.27]$ ~ & ~$\Omega_c \in [0.26,0.28]$ ~ & ~$\Omega_c \in [0.25,0.29]$~ \\ \hline
{~$\Omega_f \in [0.00,0.25]$~} & $4.534\cdot10^{-71}$ & $2.655\cdot 10^{-71}$ & $7.797\cdot10^{-72}$ & $2.258\cdot 10^{-71}$  \\ \cline{1-1}
{~$\Omega_f \in [0.25,0.50]$~} & $2.060\cdot10^{-72}$ & $5.970\cdot10^{-72}$ & $1.242\cdot10^{-71}$ & $7.289\cdot10^{-72}$     \\ \cline{1-1}
{~$\Omega_f \in [0.50,0.75]$~} & $4.312\cdot10^{-81}$ & $1.667\cdot10^{-76}$ & $3.760\cdot10^{-74}$ & $4.681\cdot10^{-74}$  \\ \cline{1-1}
{~$\Omega_f \in [0.75,1.00]$~} & $5.250\cdot10^{-100}$ & $6.557\cdot10^{-90}$ & $8.821\cdot10^{-83}$  & $2.975\cdot10^{-79}$  \\  \hline
\end{tabular}\label{tab:evidence_union}
}

\subtable[\ ESSENCE]{
\begin{tabular}{|c|c|c|c|c|}
\hline
\backslashbox{\textbf{Prior on $\Omega_f$}}{\textbf{Prior on $\Omega_c$}}& ~$\Omega_c \in [0.24,0.26]$ ~ & ~$\Omega_c \in [0.23,0.27]$ ~ & ~$\Omega_c \in [0.26,0.28]$ ~ & ~$\Omega_c \in [0.25,0.29]$~ \\ \hline
{~$\Omega_f \in [0.00,0.25]$~} & $2.052\cdot10^{-45}$ & $1.194\cdot 10^{-45}$ & $3.359\cdot10^{-46}$ & $1.003\cdot10^{-45}$    \\ \cline{1-1}
{~$\Omega_f \in [0.25,0.50]$ ~} & $1.750\cdot10^{-46}$ & $4.628\cdot10^{-46}$ & $9.751\cdot10^{-46}$ & $5.798\cdot10^{-46}$   \\ \cline{1-1}
{~$\Omega_f \in [0.50,0.75]$ ~} & $3.435\cdot10^{-54}$ & $6.867\cdot10^{-50}$ & $9.742\cdot10^{-48}$ & $1.058\cdot10^{-47}$   \\ \cline{1-1}
{~$\Omega_f \in [0.75,1.00]$ ~} & $8.433\cdot10^{-72}$ & $4.257\cdot10^{-62}$ & $2.501\cdot10^{-55}$ & $4.115\cdot10^{-52}$   \\ \hline
\end{tabular}\label{tab:evidence_essence}}

\end{center}
\caption{\label{tab:evidences} Bayesian evidences for our unified dark energy DBI model from the combination of SN+CMB+BAO observations for two different SN compilation samples.}
\end{table*}

\section{Model kinematics}
As this is a new model it is worth examining it from different perspectives, the kinematic one being a specially relevant one.
We investigate the redshift dependence of the effective equation of state parameter, $w(z)$, and derived quantities such as the acceleration parameter, $q(z)$, or the transition redshift, $z_t$. In order to obtain the  behavior of $w(z)$, we use the expression that relates it with the Friedman equation \cite{Saini2000,Nesseris2004}
\be
\displaystyle{w(z)}=\frac{\displaystyle{\frac{2}{3}\frac{d\ln{H}}{dz}(1+z)-1}}{\displaystyle{1-\left(\frac{H_0}{H}\right)^2\Omega_{c}(1+z)^3}}. 
\ee

In our model it takes the form
\begin{widetext}
\begin{equation}
\displaystyle{w(z)=\frac{\left({\Omega_r} (1+z)^4-3{\Omega_\Lambda}\right) \sqrt{{\Omega_c}^2 (1+z)^6+{\Omega_f}^2}-3 {\Omega_f}^2}{3 \sqrt{{\Omega_c}^2 (1+z)^6+{\Omega_f}^2} \left({\Omega_r}
   (1+z)^4-{\Omega_c} (1+z)^3+{\Omega_\Lambda}+\sqrt{{\Omega_c}^2 (1+z)^6+{\Omega_f}^2}\right)}}.
\end{equation}
\end{widetext}

\begin{figure*}
\subfigure[\ UNION]{
\includegraphics[width=0.4\textwidth]{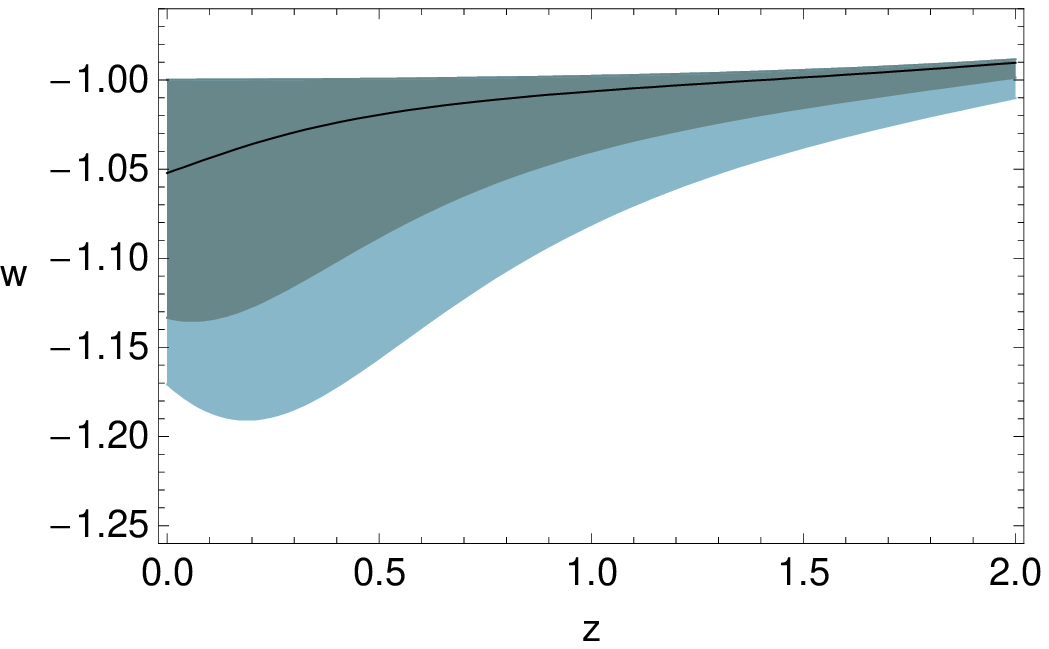}
\label{fig:w(z)union}}
\subfigure[\ ESSENCE]{
\includegraphics[width=0.4\textwidth]{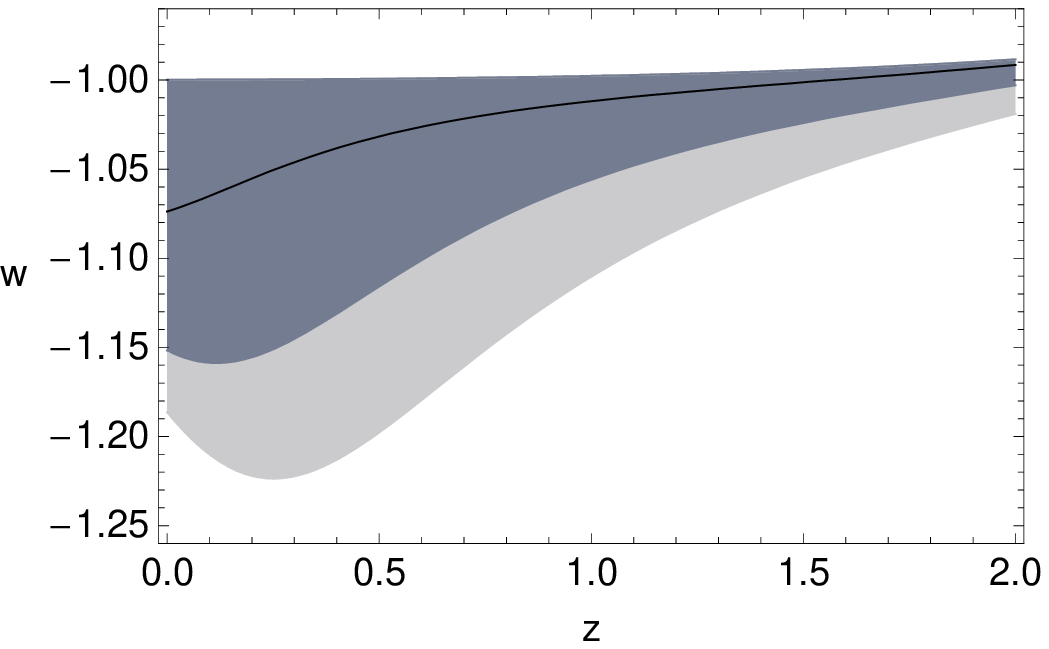}
\label{fig:w(z)essence}}
\caption{Variation of the equation of state parameter with the redshift for two different SN compilations.\label{fig:w(z)}}
\end{figure*}

\begin{figure*}
\subfigure[\ UNION]{
\includegraphics[width=0.4\textwidth]{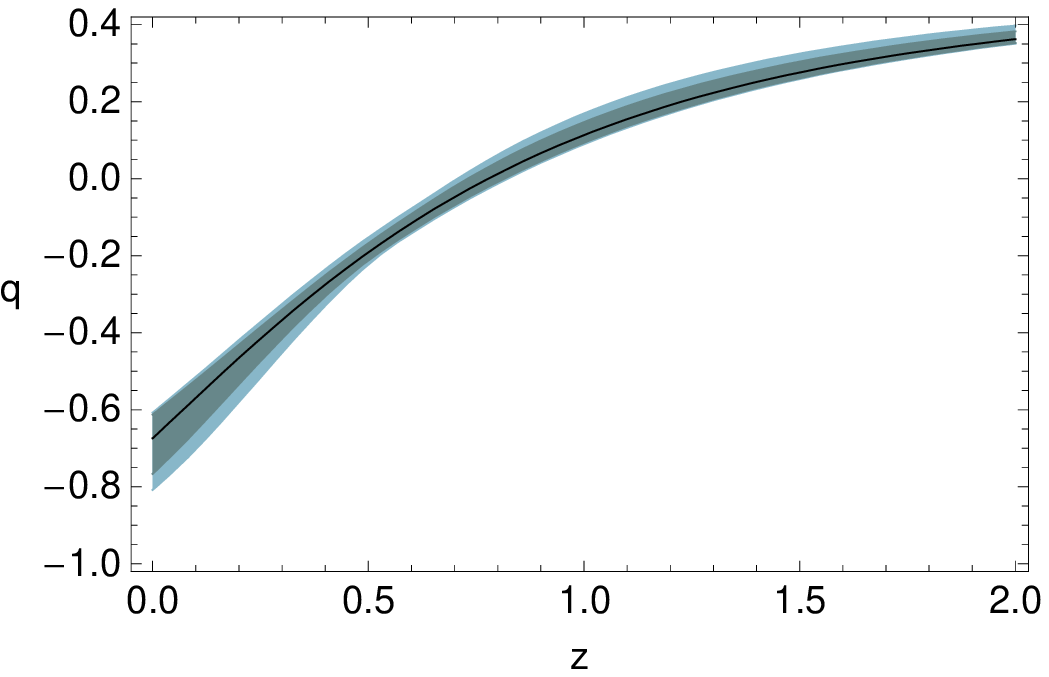}
\label{fig:qunion}}
\subfigure[\ ESSENCE]{
\includegraphics[width=0.4\textwidth]{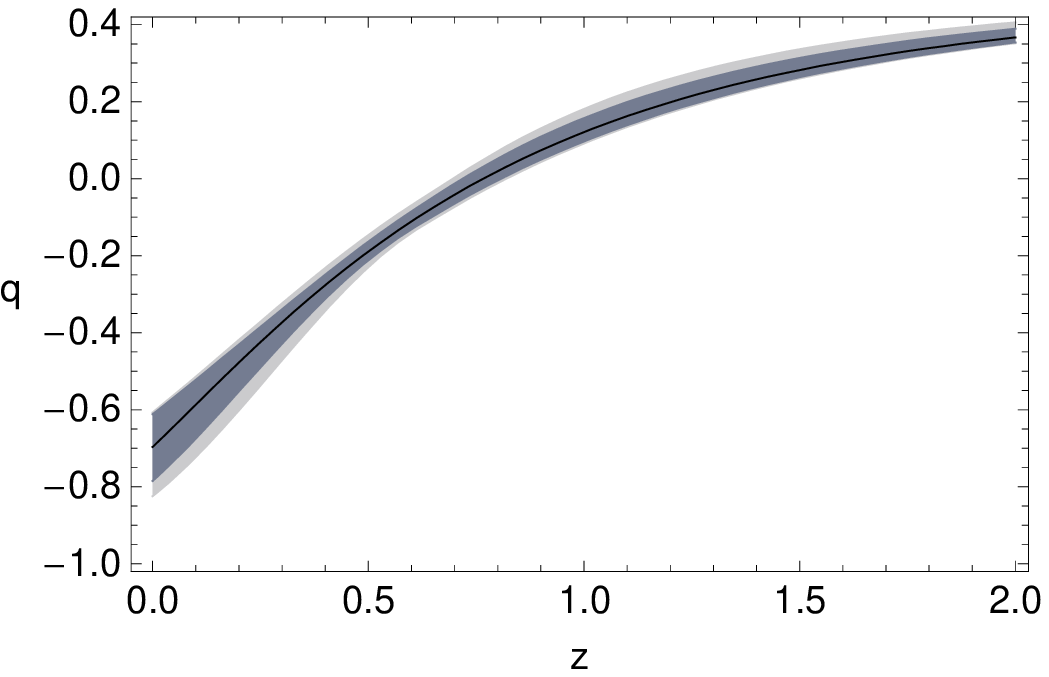}
\label{fig:qessence}}
\caption{Variation of the acceleration parameter with the redshift for two different SN compilations.\label{fig:q}}
\end{figure*}

Analyzing the dependence of the equation of state parameter with the redshift we can see that the current observational data in the context of our model restricts it to be smaller than -1, $w(z)\leq-1$ with $\left.\frac{dw}{dz}\right|_{z=0}>0$ for the current time. More precisely, we obtain $w(z=0)=-1.052^{+0.052}_{-0.081}$, $\left.\frac{dw}{dz}\right|_{z=0}=0.081^{+0.081}_{-0.140}$ with the UNION sample and $w(z=0)=-1.074^{+0.074}_{-0.078}$,$\left.\frac{dw}{dz}\right|_{z=0}=0.074^{+0.074}_{-0.199}$ with the ESSENCE sample, with the uncertainties corresponding  to the 68.30\% interval of confidence.

In addition, we study the acceleration parameter, $q(z)$ which for instance can be expressed as
\be
q(z)=\frac{3}{2}\left(1-\frac{\Omega_c\left(1+z\right)^3}{H^2}\right)w(z)+\frac{1}{2}.
\label{q(z)}
\ee
Figs. \ref{fig:w(z)} and \ref{fig:q} depict the shape of the equation of state and acceleration parameters with their corresponding 68.30\% and 95.45\% errors. From the analysis of the acceleration parameter we gather that there is a strong evidence of the transition from a deceleration to an acceleration stage. For a better insight on this matter, we have inferred the redshift at which the transition happens in different ways.

Since we have an expression for $q(z)$,
\begin{widetext}
\be
q(z)=\frac{\displaystyle{(1+z)\left(4\Omega_r(1+z)^3+\frac{3\Omega_r^2(1+z)^5}{\sqrt{\Omega_f^2+\Omega_r^2(1+z)^6}}\right)}}{\displaystyle{2\left(1-\sqrt{\Omega_c^2+\Omega_f^2}-\Omega_r+\Omega_r(1+z)^4+\sqrt{\Omega_f^2+\Omega_c^2(1+z)^6}\right)}}-1
\label{eq:q(z)}
\ee
\end{widetext}
we can compute $q(z_t)=0$ to obtain the transition redshift, $z_t$. For the ESSENCE compilation of SNIa we have $z_t=0.766^{+0.041}_{-0.047}$, and for the UNION sample $z_t=0.778^{+0.036}_{-0.048}$.

In \cite{Riess2004} another approach to obtain it was proposed. It involves expanding the acceleration parameter, $q(z)$, into two terms:
\be
q(z)=q_0+\left.z\frac{dq}{dz}\right|_{z=0}.
\ee

Under this definition, we get with the UNION sample $z_t=0.649^{+0.096}_{-0.079}$ and with the ESSENCE sample $z_t=0.650^{+0.139}_{-0.078}$. These results are in good agreement with the results obtained in \cite{Wang2006,Riess2004,Gong2007}. 

Yet another parametrization was considered in \cite{Xu2007,Cunha2008}:
\be
q(z)=q_0+q_1\frac{z}{1+z},
\ee
where $q_0=q(z=0)$ is the value of the deceleration parameter at present and $q_1$ is the parameter that contains the correction in the distant past ($q(z)=q_0+q_1$ for $z\gg0$). With this parametrization we get that the value of the transition redshift for ESSENCE is $z_t=0.674^{+0.092}_{-0.062}$ and for UNION $z_t=0.697^{+0.088}_{-0.085}$.

As the value of $z_t$ obtained directly by the explicit equation (\ref{eq:q(z)}) is bigger that the approximate one, we infer that the approximations are not good enough for accounting accurately for the tendency of our DBI fluid to inducing a phantom stage. The result obtained by our procedure tells us that the acceleration-deceleration transition happens before than the other definitions allow to estimate. 

\section{Conclusions}

This paper offers a new alternative to the popular models which attempt at a unification of the dark matter and dark energy components of the Universe. This new model stems from a purely kinetic DBI action, and therefore suggest that these non-canonical actions, which have been resorted to as a way to model the early acceleration in the Universe, can also serve the same purpose for the late time acceleration. At the background level the evolution of this model we put forward is like that of a cosmological scenario filled with a Chaplygin gas a cosmological constant, but this identification ceases when we explore further features of the models. To begin with this evolution is realized with a single fluid (unlike the mentioned Chaplygin gas plus cosmological constant combination) and the differences emerge 
in consequence for quantities important at the level of perturbations, specifically the difference becomes apparent in the speed of sound of the model, so this novel scenario demands an analysis of its own. We can  summarize the results as follows. At early times the divergence of the velocity perturbation is negligible, whereas the energy density perturbation is a growing one, thus signaling the initial unstable phase required for the onset of structures. At late times the velocity and energy density perturbations decouple, and the latter becomes negligible as the Universe becomes dominated by vacuum energy.  

The observational analysis suggests our model presents some attractive features which extend its value beyond the  theoretical perspective. To begin with current constraints show our model is by far better suited to the observations
that the most popular unified dark sector model: the Chaplygin gas \cite{Kamenshchik:2001cp,Bento:2002ps}. Our results also indicate a modest preference of our model as compared to the LCDM one. Perhaps the most remarkable outcome of this observational analysis is that the best fit corresponds to a phantom behaviour, i.e. the effective equation of state parameter $w_{\rm eff}$ lies at
present below the $-1$ line. It must be remember that this behaviour is achieved without actually having to resort to
a genuine phantom component, so we do not have to be concerned with the associated instabilities.

This study, which has been carried out from different relevant angles and the results achieved, convinces us that our model represents a worthy model for the unification of the dark sector, reinforces the theoretical interests of DBI models by extending the range of interest to the late Universe, and suggests the interest of exploring generalizations
of this model, probably by relaxing the assumption of a purely kinetic Lagrangian, as perhaps further degrees of freedom would allow and even better suitability to astronomical observations.  

\appendix
\section{\label{estsec} Statistics and data analysis}
In the context of a given physical model which depends on some parameters, besides fixing the ``most likely'' values of the parameters to  yield a series of available observational data, one should measure the degree of confidence in the fact that these data were generated by these parameters in an estimated interval.
\subsection{Parameter estimation}

The likelihood function, ${\cal L}(\textbf{d}\vert \btheta,{\cal M})$, is defined as the unnormalized probability density  of measuring the data $\textbf{d}=\left\{d_1,\dots, d_n\right\}$ given the model ${\cal M}$ and taking its parameters the values $\btheta=\left\{\theta_1,\dots,\theta_{\nu}\right\}$ \cite{jussi}. 

Despite our aim to keep the discussion in this section as general as possible, when we analyze particular datasets we will assume, as usual, that the measurements are normally distributed around their true value, so that
\begin{equation}\label{gauss lik}
{\cal L}(\textbf{d}\vert \btheta,{\cal M})\propto e^{-\chi^2(\btheta)/2}.
\end{equation}

The probability density function $p(\btheta \vert \textbf{d}, {\cal M})$  of the parameters to have values
$\btheta$ for the data, $\textbf{d}$, under the assumption that the true  model is ${\cal M}$ is provided by Bayes' theorem \cite{jussi}

\begin{equation}
p(\btheta \vert \textbf{d},{\cal M})= \frac{{\cal L}(\textbf{d}\vert \btheta,{\cal M})\pi(\btheta,{\cal M})}{\int{\cal L}(\textbf{d}\vert \btheta,{\cal M})\pi(\btheta,{\cal M})d\btheta},
\end{equation}
where $p(\btheta \vert \textbf{d},{\cal M})$ and $\pi(\btheta,{\cal M})$ are the posterior and prior probability density functions (pdf) respectively \cite{jussi, Trotta2004,Trotta2007, cousins, holmes}. The prior pdf encodes all previous knowledge about the parameters before the observational data have been collected. It can be regarded as a subjective procedure, but its use is compulsory in  the Bayesian framework, which is the approach used in theoretical frameworks
where only one particular realization of the measurement is available.

Parameter estimation in the Bayesian framework is based on maximizing the posterior pdf $p(\btheta \vert \textbf{d},{\cal M})$, whereas in a ``strict''   frequentist approach  one just maximizes $ {\cal L}(\textbf{d} \vert \btheta,{\cal M})$. When one uses flat priors in the Bayesian approach then the same conclusions are drawn from both approaches and then the difference turns to be conceptual only  \cite{Trotta2004,Trotta2007,trotta3}. If the measured observables are independent form each other and Gaussian distributed around their true value, $\textbf{d}(\btheta)$, with a covariance matrix, $\matrix{C}$, given by the experimental errors, maximizing ${\cal L}$ is equivalent to minimizing the chi-square function

\be
\chi^2(\btheta)\equiv\left(\textbf{d}^{obs}-\textbf{d}(\btheta)\right)\matrix{C}^{-1}\left(\textbf{d}^{obs}-\textbf{d}(\btheta)\right)^{T}
\ee

and for uncorrelated data $\matrix{C}_{ij}=\delta_{ij}\sigma_i^2$,
\be
\chi^2(\btheta)\equiv\sum_{i=1}^n\left(\frac{\textbf{d}^{obs}-\textbf{d}(\btheta)}{\bsigma^{obs}_i}\right)^2.
\ee

The second step toward constraining parameters satisfactorily is to construct credible intervals \cite{Trotta2004} which measure the degree of confidence that a certain data was generated by parameters belonging to the estimated interval.

In the Bayesian approach, the credible intervals are drawn around the maximum likelihood point, which gives the best fit parameters. After obtaining it by the minimization of the $\chi^2(\btheta)$, the boundaries of the region containing $100n\%$ of likelihood are determined as the values of the parameters for which $\chi^2$ has increased by a certain quantity
\be
\chi^2-\chi^2_{min}=\Delta_{\nu,n}
\ee

with

\be
n=1-\frac{\displaystyle{\int^\infty_{\Delta_{\frac{\nu,n}{2}}}t^{\frac{\nu}{2}-1}e^{-t}dt}}{\displaystyle{\int^\infty_{0}t^{\frac{\nu}{2}-1}e{-t}dt}}=1-\frac{\displaystyle{\Gamma\left(\frac{\nu}{2},\frac{\Delta_{\nu,k}}{2}\right)}}{\displaystyle{\Gamma\left(\frac{\nu}{2}\right)}}
\ee
where $\Gamma\left(\frac{\nu}{2},\frac{\Delta_{\nu,k}}{2}\right)$ is the incomplete $\Gamma$ function \cite{Lazkoz2005}, \cite{Press1992}.

The $1\sigma$ and $2\sigma$ errors of the parameter $\theta_i$ are given by the $68.30\%$ and $95.45\%$ credible interval contours, respectively. The upper limit is the maximum value of the contour and the lower one the minimum one.

\subsection{Bayesian evidence}
In Bayes' approach the evidence is an employed tool which informs about how well the parameters of the model fit the data, after doing an averaging over all the parameter values that were theoretically plausible before the measurement ever took place \cite{liddle}.

Then the Bayes' evidence is calculated as the average likelihood of the model over its prior parameter space, 
\begin{equation}
{\cal E(M)}=\int\pi(\btheta,{\cal M}){\cal L}(\textbf{d} \vert \btheta ,{\cal M}) d\btheta,
\end{equation}
where $\pi(\btheta,{\cal M})$ is the model's prior on the set of parameters normalized to unity (i.e.
($\int \pi(\btheta,{\cal M})d\btheta=1$.)
The most common choice is the top hat prior, $\pi(\btheta,{\cal M})=1/V$ with $V=\prod_{\alpha=1}^{\nu}\left(\btheta_{\alpha,max} -\btheta_{\alpha,min}\right)$. In that case one rewrites  Bayes evidence as
\begin{equation}
{\cal E(M)}=\frac{1}{V}\int_V{{\cal L}\left(\btheta\right)d\btheta}.
\end{equation}

One important and unavoidable inconvenient of the use of the evidence is its dependence on the prior ranges chosen for parameters. In this way we have computed the evidences corresponding to different prior ranges from comparison in order to find the most suitable one for our model, see Tab.\ref{tab:evidences}.

Once we arrived at this point, a remark is required. The usual situation in cosmology is that one has more than one set of statistically independent observational data, $\{\textbf{d}^{(1)}\},\dots \{\textbf{d}^{(m)}\}$ in order to constrain the parameters $\btheta$; in that case, one can resort to the joint probability density function 
\begin{eqnarray}
p(\btheta \vert \textbf{d}^{(1)}\cap\dots\cap \textbf{d}^{(m)}, {\cal M})=\nonumber\\p(\btheta \vert \textbf{d}^{(1)}, {\cal M})\times
\dots\times
p(\btheta \vert \textbf{d}^{(m)}, {\cal M}).
\end{eqnarray}
With the latest rule one can generalize conveniently the whole discussion above to the situation with more than one dataset.

\section{Error propagation in derived quantities}
 In our results, the parameters have not symmetric errors. Then we can not use the standard error propagation formula and we have to perform a modification in order to account for these non-gaussianities, \cite{Lazkoz2007}. In our case, the constraints on the parameters are given in the form, ${\theta_i}^{+\delta\theta_{i,u}}_{-\delta\theta_{i,d}}$, where $\delta\theta_{i,u}$ and $\delta\theta_{i,d}$ are positive quantities.
 
The estimated error in a quantity depending on them, $f(\btheta)$, will be given by an upper limit

\begin{equation}
\Delta f_u= \sqrt{\sum_{i=1}^n\left({\rm max}\left(\Delta f_{iu},-\Delta f_{il}\right)\right)^2}
\end{equation}

and a lower one

\begin{equation}
\Delta f_l= \sqrt{\sum_{i=1}^n\left({\rm min}\left(\Delta f_{iu},-\Delta f_{il}\right)\right)^2},
\end{equation}

where
\begin{equation}
 \Delta f_{iu}=f(\dots\theta_{(i-1)},\theta_{i}+\Delta\theta_{iu},
\theta_{(i+1)},\dots)-f(\btheta)
\end{equation}
\begin{equation}
 \Delta f_{il}=f(\dots\theta_{(i-1)},\theta_{i}-\Delta\theta_{il},
\theta_{(i+1)},\dots)-f(\btheta).
\end{equation}

This error estimation is based on finite differences, however it can be refined if the errors are enough small, i.e. $\Delta\theta_{i,u}=\delta\theta_{i,u}$ and $\Delta\theta_{i,l}=\delta\theta_{i,l}$. In that case one can write

\begin{equation}
\Delta f_u\simeq\delta f_u= \sqrt{\sum_{i=1}^n\left({\rm max}\left(\frac{\partial f}{\partial \theta_i}\delta\theta_{iu},-\frac{\partial f}{\partial \theta_i}\delta\theta_{il}
\right)\right)^2}
\end{equation}
and 
\begin{equation}
\Delta f_l\simeq\delta f_l= \sqrt{\sum_{i=1}^n\left({\rm min}\left(\frac{\partial f}{\partial \theta_i}\delta\theta_{iu},-\frac{\partial f}{\partial \theta_i}\delta\theta_{il}
\right)\right)^2}.
\end{equation}

In Gaussian situations, where $\Delta\theta_{i,u}=\Delta\theta_{i,l}=\Delta\theta_i$, one gets the standard error propagation formula and $\Delta f_u=\Delta f_l$.

\section{Observational tests}

\subsection{CMB test}
The peaks and troughs of acoustic oscillations are sensitive to the distance to the decoupling epoch. Therefore CMB provides a measure of the ratio of angular diameter distances to the decoupling epoch divided by the sound horizon size at this time, $D_A(z_\ast)/r_s(r_\ast)$. Since we have assumed a flat universe, instead of $D_A(z)$, we can use  the comoving distance
\be
D_c(z)=c\int_0^z\frac{dz'}{H(z')}.
\ee
We can determine the ratio $D_c(z_\ast)/r(z_\ast)$ by the ``acoustic scale", $l_A$,

\be
l_A(z_\ast)\equiv\frac{\pi D_c(z_\ast)}{r_s(z_\ast)}.
\ee

In this case, we use the fitting function of $z_\ast$ proposed in \cite{Hu1996}

\be
z_\ast=1048\left[1+0.00124\left(\Omega_b h^2\right)^{-0.738}\right]\left[1+g_1\left(\Omega_ch^2\right)^{g_2}\right]
\ee
with
\be
g_1=\frac{0.0783\left(\Omega_bh^2\right)^{-0.238}}{1+39.5\left(\Omega_bh^2\right)^{0.763}}
\ee

\be
g_2=\frac{0.560}{1+21.1\left(\Omega_bh^2\right)^{1.81}}.
\ee
and the comoving sound horizon given by
\be
r_s(z)=\frac{c}{\sqrt{3}}\int^{1/\left(1+z\right)}_0\frac{da}{a^2H(a)\sqrt{1+\frac{3\Omega_b}{4\Omega_\gamma}a}}
\ee
with $\Omega_\gamma=2.469\cdot10^{-5}h^{-2}$, $c=2.9979\cdot10^5$ (for $T_{CMB}=2.725K$) and $h=0.72$ \cite{Komatsu2008}.

CMB also gives a measure of the ``shift parameter", $R(z)$, which is related to $D_c$ by \cite{Bond1997}

\be
R(z_\ast)\equiv\sqrt{\Omega_c H_0^2}D_c(z_\ast).
\ee

Constructing a vector containing these quantities, $\textbf{v}=(l_A, R, z_\ast)$, and using the maximum likelihood values of 5-year WMAP \cite{Komatsu2008}, $\textbf{v}^{\rm CMB}=(302.10, 1.710, 1090.04)$ one can compute the corresponding $\chi^2$,
\be
\chi^2_{\rm CMB}=(v_i-v^{\rm CMB}_i)(\matrix{C}^{-1})_{ij}^{\rm CMB}(v_j-v^{\rm CMB}_j)^{T}
\ee
where $\left( \matrix{C^{-1}}\right)^{CMB}$ is the inverse covariant matrix of the data.

This derivation of WMAP distance priors restricts the models to test because it requires that we assume an certain cosmological scenario \cite{Komatsu2008}.
\subsection{BAO test}

There is a dependence between the peak position of the Baryon Acoustic Oscillations (BAO) and the ratio of $D_V(z)$ to the sound horizon size at drag epoch, $r_s\left(z_{\rm{drag}}\right)$, at which the baryons were liberated from photons.
$D_V(z)$ is a effective distance measure related to the comoving distance

\be
D_V(z)\equiv\left[D_c^2(z)\frac{cz}{H(z)}\right]^{1/3}.
\ee.

In order to calculate the drag epoch, we use the formula put forward in \cite{Eisenstein1997}

\be
z_{\rm{drag}}=\frac{1291\left(\Omega_ch^2\right)^{0.251}}{1+0.659\left(\Omega_ch^2\right)^{0.828}}\left[1+b_1\left(\Omega_bh^2\right)^{b_2}\right]
\ee
where

\be
b_1=0.313\left(\Omega_ch^2\right)^{-0.419}\left[1+0.607\left(\Omega_ch^2\right)^{0.674}\right]
\ee
and

\be
b_2=0.238\left(\Omega_ch^2\right)^{0.223}.
\ee

Now, taking into account the Gaussian priors at $z=0.2$ and $0.35$  from BAO data appearing in \cite{Percival2007}, we calculate
\be
\chi^2_{\rm BAO}=(v_i-v^{\rm BAO}_i)(\matrix{C}^{-1})^{\rm BAO}_{ij}(v_j-v^{\rm BAO}_j)^{T}
\ee

with $\textbf{v}=\left\{\frac{r_s(z_{\rm{drag}})}{D_V(0.2)},\frac{r_s(z_{\rm{drag}})}{D_V(0.35)}\right\}$ and $\textbf{v}^{\rm BAO}=\left(0.1980,0.1904\right)$.
\subsection{Type Ia Supernovae}
The reduced observational data usually reports values of the distance modulus
\be
\mu_{th}(z_i)=5\log_{10}\left(d_L\left(z;\btheta\right)\right) + \mu_0
\ee
with the dimensionless luminosity distance.

\be
d_L\left(z;\btheta\right)=(1+z)\int^z_0\frac{H_0dz}{H\left(z;H_0,\btheta\right)}.
\ee

Then the $\chi^2$ function to minimize takes the form

\be
\chi^2_{\rm SN}(\mu_0,\btheta)=\sum_{j=1}\frac{(\mu_{\rm th}(z_j;\mu_0,\btheta)-\mu_{\rm obs}(z_j))^2}{\sigma_{\mu,j}^2},
\ee

where $\sigma_{\mu,j}$ are the measurement variances. But there is a nuisance parameter, $\mu_0$, which makes the computation of $\chi^2$ more intensive as this parameter is marginalized over.
Often is used an alternative to marginalize it which consists in maximizing the likelihood by minimizing $\chi^2$ with respect to $\mu_0$ \cite{Elgaroy2006}.
Then one can rewrite the $\chi^2$ as

\begin{equation}
\chi^2_{\rm SN}(\btheta)=c_1-2c_2\mu_0+c_3\mu_0^2
\end{equation}
being
\begin{eqnarray}
&&c_1=\sum_{j=1}\frac{\left(\mu_{\rm obs}(z_j)-5\log_{10}d_L\left(z_j;\btheta\right)\right)^2}{\sigma_{\mu,j}^2}\quad\\
&&c_2=\sum_{j=1}\frac{\mu_{\rm obs}(z_j)-5\log_{10}d_L\left(z_j;\btheta\right)}{\sigma_{\mu,j}^2}\quad\\
&&c_3=\sum_{j=1}\frac{1}{\sigma_{\mu,j}^2}\quad.
\end{eqnarray}

The minimization over $\mu_0$ gives $\mu_0=c_2/c_3$. So the $\chi^2$ function takes the form
\begin{equation}
\tilde\chi^2_{\rm SN}(\btheta)=c_1-\frac{c_2^2}{c_3}. 
\end{equation}

The difference with respect to marginalization over $\mu_0$ is negligible in our results \cite{Nesseris2004}.

\begin{acknowledgments}
We thank A. D\'\i ez-Tejedor for comments.
L.P.C. is partially supported by the University of
Buenos Aires for partial support under project X224,
and the Consejo Nacional de Investigaciones Cient\'\i ficas
y T\'ecnicas under project 5169. R.L. and I.S. are supported by the former Spanish Ministry of Science and Innovation
through research grant FIS2007-61800. R.L. has also the support of  the University of the Basque Country through research grant GIU06/37.
\end{acknowledgments}

\bibliographystyle{h-physrev3}
\bibliography{tesis,DBI_introduction}
\end{document}